\documentstyle[prl,aps,epsf]{revtex} 
\begin{document}
\draft

\twocolumn[
\hsize\textwidth\columnwidth\hsize\csname@twocolumnfalse\endcsname

\title{Theory of ac Josephson effect and noise in 
superconducting constrictions } 
\author{Dmitri V. Averin} 
\address{Department of Physics and Astronomy, SUNY at Stony Brook, 
Stony Brook, New York, 11794, USA }

\maketitle

\vspace*{3ex}

]

\vspace*{-1ex}

\section{Introduction}

In contrast to point contacts between normal metals, the
superconducting point contacts have very non-trivial transport
properties even under the simplest bias conditions of fixed dc bias
voltage $V$ across the contact.  The origin of this complexity is the
oscillating Josephson current induced in the point contact by
non-vanishing bias voltage which  makes electron motion through the
contact inelastic.  From the microscopic point of view, this type 
of inelastic electron dynamics can be understood in terms of cycles of 
Andreev reflections at the two interfaces between the contact and
superconducting electrodes \cite{kbt}. Electron emerging from, say,
the left electrode with energy $\varepsilon$ relative to the Fermi level 
of this electrode is accelerated by the applied voltage $V$ and reaches 
the right electrode having the energy $\varepsilon +eV$.  It is then 
Andreev-reflected back as a hole which is again {\em accelerated} when 
crossing the contact backward. After being Andreev-reflected from the 
left electrode this hole produces an electron with the energy 
$\varepsilon +2eV$ (Fig.\ 1a). Because of this process
of multiple Andreev reflections (MAR) electron with energy
$\varepsilon$ incident on the contact can absorb or emit some number
$n$ of quanta $2eV$ of the Josephson oscillations and emerge from the
contact area with the energy $\varepsilon +2neV$.

Phenomenologically, the cycles of MAR manifest themselves in the
so-called ``subharmonic gap structure'' (SGS) in the current-voltage
characteristics of the superconducting contacts:  current
singularities at voltages $V_n=2\Delta/en$, $n=2,3,...$, where $\Delta$
is the superconducting energy gap of the contact electrodes.  Since
the cycle of two Andreev reflections requires that both electron and
hole traverse the contact, the probability of this process  is proportional 
to $D^2$ in contacts with small transparency $D$.  This means that
in junctions with the low-transparent barriers, the MAR current is
much smaller than the regular quasiparticle current or current of Cooper 
pairs which both scale as $D$ with junction transparency. There is, however, 
a very large variety of high-transparency Josephson junctions, where 
multiple Andreev reflections and associated subharmonic gap structure are 
important. Possible implementations of the high-transparency junctions 
include tunnel junctions with high critical current density \cite{akl}, 
semiconductor/superconductor heterostructures \cite{tak,mooij}, 
controllable break junctions \cite{post,oub,sacley}, disoredered
superconductor/semiconductor junctions \cite{ova,lind}, short metallic 
SNS junctions \cite{ohta}.

\begin{figure}[htb]
\setlength{\unitlength}{1.0in}
\begin{picture}(2.0,2.8)(0.1,0) 
\put(0.15,-0.5){\epsfxsize=3.0in\epsfysize=3.0in \epsfbox{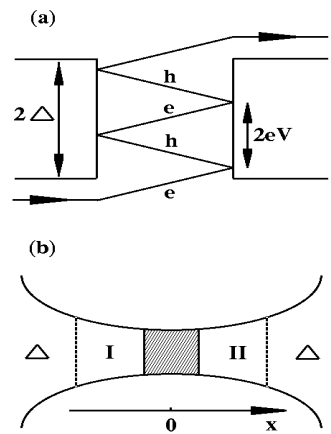}}
\end{picture}
\caption{(a) Schematic energy diagram of the multiple Andreev 
reflections process in a ballistic constriction between two 
superconductors with energy gap $\Delta$. Both electrons and holes 
are accelerated by the applied bias voltage $V$ and gain energy $eV$ 
crossing the constriction. (b) Geometry of the constriction. The shaded 
area corresponds to the scattering region. The pair potential $\Delta(x)$ 
can be neglected in regions I and II. }
\label{f1} 
\end{figure}

The subharmonic gap structure was extensively studied in experiments 
done in the 60's \cite{61,62,63,64,65}, and was interpreted in terms
of the multi-particle tunneling \cite{mpt}.  However, the theory
\cite{mpt} of multi-particle tunneling is valid only in the limit of 
low junction transparency,
$D\rightarrow 0$, and can not account quantitatively for all
features of the SGS.  A more detailed understanding of the SGS was
developed later \cite{kbt,zai1,zai2,arn} based on the idea of MAR.
Nevertheless, quantitative dependence of the SGS on the contact transparency
$D$ was still not fully understood at that time. During the last few
years, considerable progress has been made in quantitative description
of MAR in short constrictions (shorter that the coherence length $\xi$
of the superconducting electrodes of the constriction) with arbitrary
transparency \cite{sw,we1,gz,we2,we3}. The aim of this Chapter is to 
review these recent
developments.  Section 2 presents the scattering approach to MAR in
the multi-mode constrictions, in particular, in short disordered SNS 
junctions. In Section 3 we discuss the Green's function approach to 
description of MAR which is needed to account for the effects of 
non-trivial microscopic structure of the constriction electrodes. 
In Section 4 the noise properties of the ballistic constrictions 
are calculated starting from the results obtained in Section 3. 
The Conclusion summarizes 
the results and unsolved problems of the theory of ac Josephson 
effect in high-transparency Josephson junctions.

\section{Multi-mode constrictions}

The main simplification brought about by the condition $d\ll \xi$,
where $d$ is the characteristic dimensions of the constriction, is the
possibility of neglecting, first, all properties of the constriction 
besides its normal-state scattering matrix $S$, and second, all 
deviations from equilibrium in the electrodes of the constriction 
\cite{ko1,zai1}. 
These two factors allow to avoid the self-consistent determination 
of both the pair potential and electrostatic potential. Electron 
transport through the constriction is determined then by the 
interplay of scattering in the constriction (described by $S$), 
acceleration of quasiparticles in the constriction by the bias 
voltage $V$, and Andreev reflection with the amplitude $a(\varepsilon)$ 
at the two interfaces with the superconducting electrodes in equilibrium,   
\begin{equation}
a(\varepsilon)=\frac{1}{\Delta} \left\{ \begin{array}{ll} 
\varepsilon- \mbox{sign}(\varepsilon) (\varepsilon^2- \Delta^2)^{1/2}\, , 
\;\;\; & \mid \varepsilon \mid > \Delta\, , \\
\varepsilon-i(\Delta^2 -\varepsilon^2)^{1/2}\, , \;\;\; & \mid 
\varepsilon \mid < \Delta\, . \end{array} \right. 
\label{aa} \end{equation}  
Here $\varepsilon$ is the quasiparticle energy relative to the Fermi 
level of the electrode, and $\Delta$ is the equilibrium energy gap of 
the electrodes.

To find the transport properties of the constriction quantitatively, 
we generalize \cite{we1,dis} to non-vanishing bias voltages $V$ the 
scattering approach for Bogolyubov-de Gennes equations \cite{fur,been,bag} 
that describes 
dc Josephson effect at $V=0$. In accordance with the qualitative picture 
of MAR discussed in the Introduction, electron with energy $\varepsilon$   
incident on the constriction generates electron and hole states 
at energies $\varepsilon +2neV$ with arbitrary $n$. Thus, the electron 
and hole wavefunctions in regions I and II of the constriction (Fig.\ 1b) 
can be written as follows: 
\begin{equation} 
\mbox{(I)} \; \begin{array}{l} 
\psi_{el} =\sum_{n}[(a_{2n}A_n+J\delta_{n0})e^{ikx}+B_ne^{-ikx} ] 
 e^{-i(\varepsilon +2neV)t/\hbar } ,  \\ 
\psi_{h} =\sum_{n}[A_ne^{ikx}+a_{2n}B_ne^{-ikx} ] 
e^{-i(\epsilon +2neV)t/\hbar } , \end{array} 
\label{I} \end{equation}
\begin{equation} 
\mbox{(II)} \; \begin{array}{l} 
\psi_{el} =\sum_{n}[C_ne^{ikx}+a_{2n+1} D_ne^{-ikx} ] 
 e^{-i(\varepsilon +(2n+1)eV)t/\hbar } ,  \\ 
\psi_{h} =\sum_{n}[a_{2n+1}C_ne^{ikx}+D_ne^{-ikx} ] 
e^{-i(\varepsilon +(2n+1)eV)t/\hbar } , \end{array} 
\label{II} \end{equation}
where $a_n\equiv a(\varepsilon+neV)$. In these equations we took into 
account that the amplitudes of electron and hole wavefunctions are 
related by the Andreev reflection, and also neglected variations of 
the quasiparticle momentum $k$ with energy assuming that the Fermi 
energy in the electrodes is much larger than $\Delta$. The 
quasiparticle energies in regions I and II are measured relative to 
the Fermi level in the left and right electrode, respectively. 
Since the constriction is assumed to support $N$ propagating transverse 
modes, all amplitudes of the electron and hole wavefunctions have 
transverse mode index $m$ not shown in eqs.\ (\ref{I}) and 
(\ref{II}), e.g., $A_n\equiv \{ A_{n,m} \}$, $m=1,...,N$.  The source 
term $J$ describes an electron generated in the $j$th transverse mode 
by a quasiparticle incident on the constriction from the left 
superconductor: \hspace*{.1em} $J(\varepsilon)=(1-\mid \! a_0 \! 
\mid ^2)^{1/2}\delta_{mj}$.  

The current in the constriction can be calculated in terms of the 
electron and hole wavefunctions in the region I or II. The contribution 
$i(t)$ to the current from the wavefunction (\ref{I}) is:
\begin{equation}
i(t)=\frac{e\hbar}{m} \mbox{Im} \mbox{Tr} (\psi_{el} \nabla  
\psi_{el}^{\dagger} -\psi_{h} \nabla  \psi_{h}^{\dagger} ) \, ,
\label{c1} \end{equation}
where Tr is taken over the transverse modes. Equations (\ref{I}) and 
(\ref{c1}) imply that the total current $I(t)$ in the constriction 
oscillates with the Josephson frequency $\omega_J =2eV/\hbar$ and can 
be expanded in the Fourier components: 
\[ I(t)=\sum_k I_k e^{ik\omega_J} . \]   
Substituting eq. (\ref{I}) into (\ref{c1}) and summing the contributions  
from quasiparticles incident both from the left and right superconductors 
at different energies $\varepsilon$ we obtain the Fourier components 
$I_k$ of the current: 
\[ I_k =-\frac{e}{\pi \hbar} \int_{-\mu-eV}^{\mu} d\epsilon \tanh 
\{ \frac{\epsilon}{2T} \} \mbox{Tr}[ (J J^{\dagger} \delta_{k0} + \] 

\vspace*{-2ex} 

\[ a_{2k}^* JA_k^{\dagger} + a_{-2k} A_{-k} J^{\dagger} + \] 

\vspace*{-2ex} 

\begin{equation} 
\left. \sum_n (1+a_{2n}a_{2(n+k)}^*) 
(A_nA_{n+k}^{\dagger}  
-B_nB_{n+k}^{\dagger} )) ] \right|_{\mu \rightarrow \infty} \, .  
\label{c2} \end{equation}   

The amplitudes $A$, $B$, $C$, $D$ of electron and hole wavefunctions 
(\ref{I}) and (\ref{II}) are related by the matrix $S$ of scattering 
in the constriction. Taking into account that the scattering matrix 
for the holes is the time-reversal conjugate of electron scattering 
matrix $S$ we can write:   
\begin{equation} 
\left( \begin{array}{c} B_n \\ C_n \end{array} \right)  = S 
\left( \begin{array}{c} a_{2n}A_n +J\delta_{n0} \\ 
a_{2n+1}D_n \end{array} \right) \, , 
\label{c3} \end{equation} 
\begin{equation} 
\left( \begin{array}{c} A_n \\ D_{n-1} \end{array} \right) = 
S^* \left( \begin{array}{c} a_{2n}B_n \\ 
a_{2n-1}C_{n-1} \end{array} \right) \, , 
\label{c4}  \end{equation} 
The scattering matrix $S$ is a unitary and symmetric matrix $2N 
\times 2N$ and can be written in terms of reflection and transmission 
$N\times N$ matrices $r$, $t$:  
\begin{equation} 
S_{el}= \left( \begin{array}{cc} r & t \\ t' & r'
\end{array} \right) \, ,
\label{c5}  \end{equation} 
where $t'=t^T$, $r'= -(t^*)^{-1}r^{\dagger} t$, and $tt^{\dagger} +
rr^{\dagger} =1$.   

Eliminating $A_n$ between eq.\ (\ref{c3}) and inverse of  
eq.\ (\ref{c4}) we find the relation between the amplitudes 
$B_n$ and $D_n$. Combining this relation with the expression for $D_n$ 
in terms of $B_n$ that follows from the inverse of eq.\ (\ref{c3}) 
and eq.\ (\ref{c4}) we arrive at the following recurrence relation 
for $B_n$: 
\[ tt^{\dagger} \left( \frac{a_{2n+2}a_{2n+1}}{1-a_{2n+1}^{2}} B_{n+1} -
(\frac{a_{2n+1}^2}{1-a_{2n+1}^{2}} + \frac{a_{2n}^2}{1-a_{2n-1}
^{2}}) B_n +  \right. \] 
\vspace*{-1ex}
\begin{equation} 
\left. + \frac{a_{2n}a_{2n-1}}{1-a_{2n-1}^{2}} B_{n-1} \right) - 
 [1-a_{2n}^{2}] B_n = - rJ \delta_{n0} \, , 
\label{c6} \end{equation} 
In a similar way we obtain the recurrence relation for the amplitudes 
$A_n$:  
\[ t^*t^T \left( \frac{a_{2n+1}a_{2n}}{1-a_{2n+1}^{2}} A_{n+1} -
(\frac{a_{2n}^2}{1-a_{2n+1}^{2}} + \frac{a_{2n-1}^2}{1-a_{2n-1}
^{2}}) A_n + \right. \]
\vspace*{-1ex}
\[+ \frac{a_{2n-1}a_{2n-2}}{1-a_{2n-1}^{2}}A_{n-1} +
\left. + \frac{(a_1\delta_{n1}-a_0\delta_{n0})J}{1-a_1^2} \right) - \] 
\vspace*{-1ex}
\begin{equation} 
 - [1-a_{2n}^{2}] A_n = - a_0 J \delta_{n0} \, . 
\label{c7} \end{equation} 

Since the hermitian matrix $tt^{\dagger}$ can always be diagonalized 
by an appropriate unitary transformation $U$, the recurrence relation 
(\ref{c6}) implies that the structure of the amplitudes 
$B_n$ as vectors in the transverse-mode space is:
\begin{equation} 
B_n = U^{\dagger} f_n(D) U r J \, ,
\label{c8} \end{equation} 
where $D=U tt^{\dagger} U^{\dagger}$ is the diagonal matrix of 
transmission probabilities $D_m$, $m=1,...,N$. The functions $f_n(D)$ 
are determined by the solution of the recurrence relation (\ref{c6}) 
with the diagonalized transmission matrix $tt^{\dagger}$.   
	
Equation (\ref{c8}) shows that the contribution of the amplitudes 
$B_n$ to the currents (\ref{c2}) can be written as 
\[ \mbox{Tr} [B_n B_{n'}^{\dagger} ] = (1-\mid \! a_0 \! \mid ^2) 
\mbox{Tr}  [ f_n(D) f_{n'}^*(D) (1-D) ] \, , \] 
i.e., it can be represented as a sum of independent contributions 
from different transverse modes with the transparencies $D_m$. 
Similarly, the recurrence relation (\ref{c7}) and eq.\ (\ref{c2}) 
for the currents show that the same is true for the amplitudes $A_n$. 
Therefore, the Fourier components (\ref{c2}) of the total current 
can be written as sums of independent contributions from individual 
transverse modes: 
\begin{equation} 
I_k=\sum_m I_k(D_m) \, , 
\label{cc9} \end{equation} 
where the contribution of one (spin-degenerate) mode is:    
\[ I_k(D) =\frac{e}{\pi \hbar} \left[ eV D\delta_{k0} - \int d\epsilon 
\tanh \{ \frac{\epsilon}{2T} \} (1-|a_0|^2)(a_{2k}^*A_k^* + \right. \] 
\begin{equation} 
\left. a_{-2k} A_{-k} + \sum_n (1+a_{2n}a_{2(n+k)}^*) 
(A_n A_{n+k}^* -B_n B_{n+k}^*)) \right]  \, .  
\label{ccc9} \end{equation}
with the integral over $\varepsilon$ taken in large symmetric limits. 
The amplitudes $B_n$ in this equation are determined by the 
recurrence relation which follows directly from eq.\ (\ref{c6}):   
\[ D\frac{a_{2n+2}a_{2n+1}}{1-a_{2n+1}^{2}} B_{n+1} - 
[D(\frac{a_{2n+1}^2}{1-a_{2n+1}^{2}} + \frac{a_{2n}^2}{1-a_{2n-1}
^{2}})+1-a_{2n}^{2}] B_n + \] 
\begin{equation}
+ D\frac{a_{2n}a_{2n-1}}{1-a_{2n-1}^{2}} B_{n-1} = - R^{1/2}\delta_{n0} 
\, ,  \;\;\; R\equiv 1-D\, . 
\label{c9} \end{equation}

\begin{figure}[htb]
\setlength{\unitlength}{1.0in}
\begin{picture}(2.0,4.75) 
\put(0.25,0.15){\epsfxsize=3.0in\epsfysize=4.5in \epsfbox{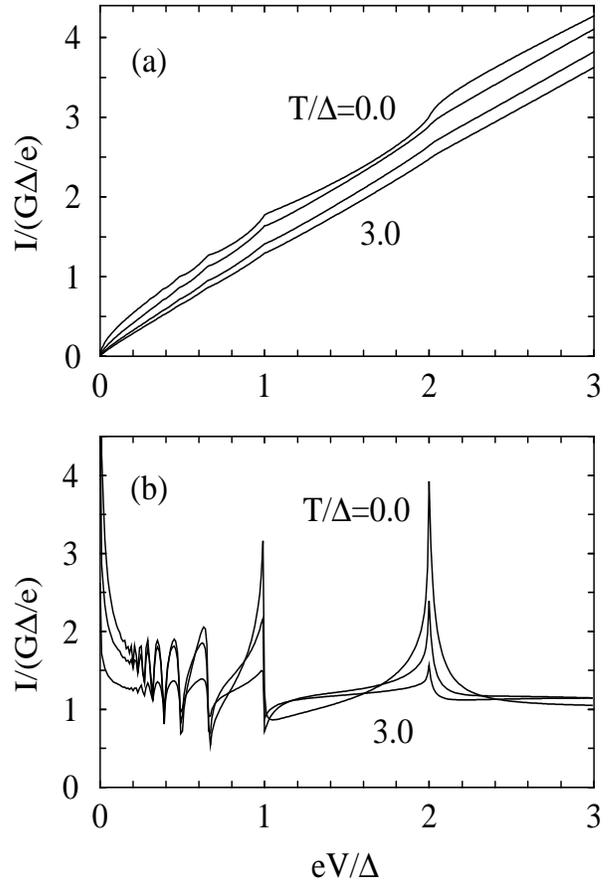}}
\end{picture}
\caption{(a) DC $IV$ characteristics and (b) differential conductance 
$dI/dV$ of a short disordered SNS junction at several temperatures:
$T/\Delta =0,\, 1,\, 2,\, 3$. (The $T=2\Delta$ curve is omitted in (b).)
The curves show the subharmonic gap 
singularities at $eV=2\Delta/n$, $n=1,2...$ associated with the multiple 
Andreev reflections. The differential conductance diverges as $V^{-1/2}$ 
at small voltages -- see text.} 
\end{figure}

Instead of using similar independent recurrence relation for $A_n$ 
that follows from eq.\ (\ref{c7}), it is more convenient to 
determine these coefficients from an equivalent relation that can be 
obtained from a single-mode version of eqs.\ (\ref{c3}) and (\ref{c4}) 
\cite{we1}:  
\begin{equation}  
A_{n+1}-a_{2n+1}a_{2n}A_n =R^{1/2}(B_{n+1}a_{2n+2}-B_na_{2n+1})  
+a_1\delta_{n0} \, ,  
\label{c10} \end{equation} 

The recurrence relations (\ref{c9}) and (\ref{c10}) can be solved 
explicitly in the case of 
perfect transmission, $D=1$. In the general case of arbitrary $D$ they 
provide an efficient algorithm for numerical evaluation of the current.  
Therefore, eqs.\ (\ref{cc9}) -- (\ref{c10}) determine completely the 
time-dependent current in a short constriction with arbitrary 
distribution of transmission probabilities. In particular, we can use 
these equations to calculate the current in a short disordered SNS 
junction with large number of transverse modes $N\gg 1$ and diffusive 
electron transport in the N region. The distribution of transmission 
probabilities is then quasicontinuous, and is characterized 
by the density function $\rho(D)$ (see, e.g., \cite{naz} and 
references therein): 
\begin{equation}  
\sum _m ... = \int_0^1 dD \rho(D) ... \, , \;\;\;\; \rho(D) = \frac{\pi 
\hbar G}{2e^2} \frac{1}{D(1-D)^{1/2} } \, , 
\label{c11} \end{equation} 
where $G$ is the normal-state conductance of the N region. This 
distribution determines 
various transport properties of disordered mesoscopic conductors, for 
instance, the magnitude of the shot-noise suppression \cite{noise1,noise2}.

Figure 2 shows the results of the numerical calculations of the dc 
current-voltage ($IV$) characteristics and differential conductance of the 
short SNS Josephson junction obtained by averaging the single-mode 
contributions to the current determined from eqs.\ (\ref{ccc9}) -- (\ref{c10})
over the distribution (\ref{c11}) . We see that the $IV$ 
characteristics has all qualitative features of the high-transparency 
Josephson junctions: subgap current singularities at $eV=2\Delta/n$ and 
excess current $I_{ex}$ at $eV\gg 2\Delta$. It is instructive to compare 
quantitatively these features to those in  the $IV$ characteristics of a 
single-mode Josephson junctions plotted in \cite{we1}. Such a comparison 
shows that the magnitude of the excess current in the SNS junction, as well  
as the overall level of current in the sub-gap region correspond 
approximately 
to a single-mode junction with large transparency $D\simeq 0.8$. At the 
same time, the subharmonic gap structure and the gap feature at 
$eV=2\Delta$ are much more pronounced than in a single-mode junction of 
this transparency. The amplitude of the oscillations of the differential 
conductance corresponds roughly to the junction with $D\simeq 0.4$ (although  
this comparison is not very accurate because of the different shapes of 
the curves). This ``discrepancy'' reflects the two-peak structure of the 
transparency distribution (\ref{c11}) of the diffusive conductor: the 
abundance of nearly ballistic modes leads to large excess and subgap 
currents, while the peak at low transparencies determines the SGS 
features. 

The main qualitative difference between the $IV$ characteristics of the  
single-mode and diffusive SNS junction lies in the low-voltage ($eV\ll 
\Delta$) behavior of the curves. As shown below, in contrast to a 
single-mode junction, the $IV$ characteristics of the SNS junction have a 
square-root singularity at low voltages which directly reflects the shape 
of the high-transparency peak in the distribution (\ref{c11}). All this 
implies that the highly nonlinear character of the $IV$ characteristics 
of the diffusive SNS junctions could serve as a direct experimental test 
of the transparency distribution (\ref{c11}) of the disordered 
mesoscopic conductor. 

Besides providing the basis for numerical evaluation of the current, 
the recurrence relations (\ref{c9}) and (\ref{c10}) can be used to 
find the current analytically at small and large voltages. At 
large voltages ($eV\gg \Delta$), the probability of MAR cycles 
decreases rapidly with the number of Andreev reflections in them. We can 
then explicitly solve the recurrence relations limiting ouselves to 
the cycles with two Andreev reflections and find the current from  
eq.\ (\ref{ccc9}). For a single mode at $T\ll \Delta$ we get the 
following high-voltage asymptote of the dc current \cite{dis}: 
\begin{equation} 
I(V) =\frac{eD}{\pi \hbar} [eV+ \frac{\Delta D}{R} (1-
\frac{D^2}{2\sqrt{R}(1+R)} \ln (\frac{1+\sqrt{R}}{1-\sqrt{R}}) ) 
-\frac{\Delta^2}{2eV} ] \, .
\label{c12} \end{equation}
Averaging this result with the distribution (\ref{c11}) we can obtain  
similar asymptote for an SNS junction: 
\begin{equation} 
I(V) = G (V+\frac{\Delta}{e} (\frac{\pi^2}{4}-1) -
\frac{\Delta^2}{2eV}) ] \, .
\label{c13} \end{equation}
The second term in this equation represents the excess current $I_{ex}$ 
and was first found in \cite{avz} by the quasiclassical Green's function 
method. It can be checked that the asymptote (\ref{c13}) agrees well 
with the numerically calculated zero-temperature $IV$ characteristic shown 
in Fig.\ 2a. The large-voltage  approximation allows us also to find the 
asymptotes of the ac components of the current \cite{dis}. 

The $IV$ characteristics of the SNS junction can also be calculated 
analytically at small voltages, $eV\ll \Delta$, using the understanding 
\cite{we1,sh2} 
that the small finite voltage $V$ drives the Landau-Zener transitions 
between the Andreev-bound states of the modes with small reflection 
coefficients $R\ll 1$. For a single mode of this type, the nonequilibrium 
voltage-induced contribution to the current at $T\ll \Delta$ is \cite{we1}:  
\begin{equation}
I(\varphi) =\frac{e\Delta}{\hbar} \left\{ \begin{array}{ll}  0 \, , 
\;\;\;\; & 0< \varphi <\pi \, , \\ 2\exp \{-\pi R\Delta/eV \} 
\sin \varphi/2 \, , \;\;\;\; & \pi < \varphi <2\pi \, ,  
\end{array} \right.    
\label{c14} \end{equation} 
where $\varphi = \varphi_0 + 2eVt/\hbar$ is the Josephson phase 
difference between the superconductors. Averaging this equation with 
the distribution (\ref{c11}) and adding the equilibrium supercurrent 
from \cite{ko2} we obtain the dynamic current-phase relation of an 
SNS Josephson junction at $eV\ll \Delta$:  
\[ I(\varphi) = \frac{G\Delta}{e} \left( \cos(\frac{\varphi}{2} ) 
\tanh^{-1}[\sin(\frac{\varphi}{2})] + \right. \] 
\vspace*{-1ex} 
\begin{equation}
\left. +\left\{ \begin{array}{ll}  
0\, ,  & 0< \varphi <\pi \, , \\ 
\pi \sqrt{eV/\Delta} \sin(\varphi/2)  \, , & \pi < \varphi <2\pi \, .  
\end{array} \right. \right)    
\label{c15}\end{equation} 
The dc current $I$ at small voltages is obtained by averaging eq.\ 
(\ref{c15}) over the phase $\varphi$:  
\begin{equation} 
I= G \sqrt{V\Delta/e} \, . 
\label{c16} \end{equation} 

This square-root behavior of the current leads to the zero-bias singularity 
of the differential conductance of the SNS junction which can be seen 
in Fig.\ 2b. Physically, this large conductance is caused by overheating 
of electrons in the N region by the MAR process.  Electrons with energies 
inside the energy gap traverse the constriction many times and as a result 
are accelerated to energies much larger the $eV$. This means that the 
effective voltage drop across the constriction is much larger than $V$, 
leading to increased conductance. This mechanism of conductance 
enhancement is qualitatively similar to the so-called ``stimulation of 
superconductivity''\cite{al} (which is one of the plausible explanations 
of the zero-bias conductance singularities \cite{kr} in long 
semiconductor Josephson junctions), although quantitatively the phenomena 
are quite different. One of the most important differences between the 
short and long SNS junctions is that the singularity (\ref{c16}) in 
short junctions should be suppressed by 
temperature simultaneously with the dc critical current, whereas in long 
junctions there is a temperature range where the zero-bias singularity 
is pronounced while the supercurrent is already negligible. 

The fact that 
the singularity (\ref{c16}) is caused by electron overheating implies 
that it should be regularized by any mechanism of inelastic scattering. 
Nevertheless, in constrictions shorter than inelastic scattering length 
$l_{in}$ in the normal metal, the square-root conductance singularity 
should be experimentally observable at low temperatures, $T\ll \Delta$, 
when the inelastic scattering in superconductors is also suppressed. 

All of the calculations presented in this Section were based on the 
assumption of ideal BCS electrodes of the constriction that are 
characterized by the Andreev reflection amplitude (\ref{aa}). This 
assumption allows to describe dynamics of Andeev reflection with the 
Bogolyubov-de Gennes (BdG) equations. 
The most noticeable drawback of this approach is the impossibility of 
incorporating inelastic scattering in the electrodes which plays an 
important role in regularizing low-voltage singularities associated 
with MAR. In the next Section we show how the BdG approach can be 
generalized to superconductors with arbitrary microscopic structure 
which in general includes inelastic processes.

\section{Superconductors with general microscopic structure}

The method that allows to discuss the inelastic effects in the 
superconducting electrodes of the constriction employs the 
quasiclassical non-equilibrium Green's functions. For the general 
introduction to this technique see, e.g., \cite{rs}. In this method, 
position dependent density of states of the superconductors is 
described with the retarded and advanced 
Green's function $G_{R,A}(x,\varepsilon,\varepsilon')$. Deep 
inside the superconducting electrodes $G_{R,A}$ reach their equilibrium 
position-independent form: 
\begin{equation} 
G_{R,A}^{(0)} (\varepsilon,\varepsilon') = \left( \begin{array}{cc} 
g_{R,A}(\varepsilon) & f_{R,A}(\varepsilon)  \\ 
-f_{R,A}(\varepsilon) & -g_{R,A}(\varepsilon)  \end{array} \right)  
\delta(\varepsilon-\varepsilon') \, ,
\label{d1} \end{equation} 
where $g$ and $f$ are the normal and anomalous Green's function of 
the superconductor which satisfy the normalization condition: 
$g^2-f^2 =1$. Retarded and advanced functions are related simply as 
\[ g_A(\epsilon)= -g_R^*(\epsilon) \, ,\;\;\; 
f_A(\epsilon)= -f_R^*(\epsilon)\, .\] 

Space dependence of $G_{R,A}$ is governed by the equation \cite{lo,zai1}:  
\begin{equation}
i \mbox{sign}(p_x)v_F \frac{\partial G_{R,A} }{\partial x}=
[H_{R,A},G_{R,A}] \, , 
\label{d2} \end{equation} 
where $v_F$ is the Fermi velocity, sign$(p_x)$ denotes two 
directions of propagation in the constriction, and $H$ is the effective 
matrix Hamiltonian of the superconductor. Since equation (\ref{d2}) 
should be satisfied also in equilibrium, $H$ as a matrix commutes with 
$G^{(0)}$. Therefore, assuming electron-hole symmetry we can express it as
\[ H_{R,A}= \Omega_{R,A} G^{(0)}_{R,A} \, , \]
where $\Omega(\varepsilon)$ is complex quasiparticle energy. 
Imaginary part of $\Omega$ comes from the imaginary part of electron 
self-energy and is positive: 
\begin{equation} 
\mbox{Im} \Omega_{R,A} > 0 \, . 
\label{omega} \end{equation} 

In order to calculate the current through the constriction we need 
to find the Green's functions inside it. Equation (\ref{d2}) shows that 
the characteristic length scale of Green's function variation is the 
coherence length $\xi=\hbar v_F/\Delta$. In our model of short constriction 
with length $d$ much smaller than $\xi$ this means that from the perspective 
of eq.\ (\ref{d2}) all points of the constriction correspond to $x=0$. 
Equation (\ref{d2}) determines the Green's functions at $x=0$ through 
the condition that corrections $\bar{G}$ to the equilibrium functions 
(\ref{d1}) should decay inside the electrodes. To see how this condition
translates into the matrix structure of $\bar{G}$ we perform a 
``rotation'' in the electron-hole space which diagonalizes $G^{(0)}$ 
(and hence $H$): 
\begin{equation}
\tilde{G}_{R,A} (\epsilon, \epsilon') \rightarrow U_{R,A} (\epsilon)  
G_{R,A} (\epsilon, \epsilon') U_{R,A}^{-1} (\epsilon')\, , 
\label{d3} \end{equation} 
where the rotation matrix is
\[ U_{R,A} = \frac{1}{\sqrt{1-a_{R,A}^{2}}} \left( \begin{array}{cc} 
1 & a_{R,A} \\ a_{R,A} & 1 \end{array} \right)   \, .  \] 
Here 
\begin{equation}
a_R (\varepsilon) =
\frac{f_R(\varepsilon)}{g_R(\varepsilon)+1} \, , \;\;\; \mbox {and} 
\;\; a_A (\varepsilon) =  a_R^* (\varepsilon)\, .  
\label{d4} \end{equation} 
We will see from the final results of this Section that $a_R$ defined 
by eq.\ (\ref{d4}) has the meaning of the amplitude of Andreev 
reflection. In particular, for the ideal BCS superconductors 
the equilibrium Green's functions are: $g_R(\varepsilon) =\varepsilon/
\delta$, $f_R(\varepsilon) =\Delta/\delta$, where $\delta \equiv 
[(\varepsilon+i0)^2 -\Delta^2]^{1/2}$, and eq.\ (\ref{d4}) reduces to 
eq.\ (\ref{aa}) of the previous Section. 

After the rotaion (\ref{d3}) equation (\ref{d2}) for the retarded 
Green's function and $p_x>0$ takes the form 
\begin{equation}
i v_F \frac{\partial \tilde{G}_R }{\partial x}= \Omega_R [\sigma_z,
\tilde{G}_R] \, , 
\label{d5} \end{equation} 
with $\sigma$'s here and below denoting Pauli matrices. Combined 
with eq.\ (\ref{omega}), this equation means that the matrix structure 
of the solutions $\tilde{G}_{R1,2}$ decaying in the first and second 
electrodes respectively, is: $\tilde{G}_{R1,2}\propto \sigma_{\pm}$, 
so that 
\[ \!\! G_{R1}(\varepsilon,\varepsilon') =  U_R^{-1}\tilde{G}_{R1} 
U_R  = q_1(\varepsilon,\varepsilon')  \left(\!\! \begin{array}{cc} 
a_R(\varepsilon') & \!\! 1 \\ -a_R(\varepsilon) a_R(\varepsilon') & 
\!\! -a_R(\varepsilon) \end{array} \!\! \right)\! , \]
\[ \!\! G_{R2}(\varepsilon,\varepsilon') =  U_R^{-1}\tilde{G}_{R2} 
U_R = q_2(\varepsilon,\varepsilon')  \left( \! \!\begin{array}{cc} 
-a_R(\varepsilon) & \! -a_R(\varepsilon) a_R(\varepsilon') \\  
1 & \! a_R(\varepsilon') \end{array} \!\! \right) \! , \] 
where $q_{1,2}$ are some functions that will be found later. 

We discuss in detail the case of a single-mode ballistic 
constriction with $D=1$. In this case the total Green's function 
should be continuous in the constriction: 
\begin{equation} 
G^{(0)}_{R1}+G_{R1}=G^{(0)}_{R2}+G_{R2} \, , 
\label{d7} \end{equation} 
where $G^{(0)}_{Rj}$ is the equilibrium Green's functions of the 
$j$th electrode. Expression (\ref{d1}) for these functions is valid 
only if the energies $\varepsilon, \varepsilon'$ are measured 
relative to the Fermi energy of the corresponding electrode. If the 
zero of energy is chosen differently, the energies in eq.\ (\ref{d1}) 
should be shifted. For instance, if  we chose the zero of energy to 
coincide with the Fermi level of the first electrode, then 
\[ G^{(0)}_{R2}= \left( \!\! \begin{array}{cc} 
g_R (\varepsilon+u) \delta(\varepsilon-\varepsilon') 
& \!\! f_R(\varepsilon+u) \delta(\varepsilon-\varepsilon'+2u) \\ 
-f_R(\varepsilon-u) \delta(\varepsilon-\varepsilon'-2u) & \!\! 
-g_R(\varepsilon-u) \delta(\varepsilon-\varepsilon') \end{array} 
\!\! \right) \! , \]
where $u\equiv eV$, while 
$G^{(0)}_{R1}$ is still given by eq.\ (\ref{d1}). Equation (\ref{d7}) 
represents then the matrix equation that allows us to determine the 
functions $q_{1,2}$. Indeed, eliminating $q_2$ from the first row of the 
matrix equation (\ref{d7}) we obtain the recurrence relation for $q_1$: 
\[ q_1(\varepsilon,\varepsilon'+2u) -q_1(\varepsilon,\varepsilon')
a_R(\varepsilon')a_R(\varepsilon'+u) =(\delta(\varepsilon-\varepsilon') 
+ \]
\vspace*{-2ex} 
\begin{equation}
+g_R(\varepsilon)) a_R(\varepsilon+u) -f_R(\varepsilon+u) 
a_R(\varepsilon') \delta(\varepsilon-\varepsilon'+2u) \, .
\label{d8} \end{equation}
Since the source terms in this relation are $\delta$-functions we can 
look for a solution in the form $q_1=\sum_n A_n(\varepsilon)
\delta(\varepsilon-\varepsilon'+2un)$. The recurrence relation for the 
coefficients $A_n$ that follows from eq.\ (\ref{d8}), complemented with 
the condition that $A_n$ should decay at $n\rightarrow \pm \infty$, can 
be solved directly. We obtain then the total Green's function in the 
constriction for $p_x>0$: 
\[ G_R = \left( \begin{array}{cc} 1 & 0  \\ -2a_R(\varepsilon) & -1 
\end{array} \right)  \delta(\varepsilon-\varepsilon') + 
2\sum_{n=1}^{\infty} \delta(\varepsilon-\varepsilon'+2un) \] 
\vspace*{-1ex}
\begin{equation}
\prod_{j=1}^{2n-1} a_R( \varepsilon+ju) \left( \begin{array}{cc} 
a_R(\varepsilon') & 1 \\ -a_R(\varepsilon) a_R(\varepsilon') & 
-a_R(\varepsilon) \end{array} \right) \, .
\label{d9} \end{equation}

In the same way we can find the Green's functions for backward 
propagation ($p_x<0$): 
\[ G_R = \left( \begin{array}{cc} 1 &  2a_R(\varepsilon) \\ 0 & -1 
\end{array} \right)  \delta(\varepsilon-\varepsilon') + 
2\sum_{n=1}^{\infty} \delta(\varepsilon-\varepsilon'-2un) \]
\vspace*{-1ex}
\begin{equation}
\prod_{j=1}^{2n-1} a_R( \varepsilon-ju) \left( \begin{array}{cc} 
a_R(\varepsilon) & a_R(\varepsilon) a_R(\varepsilon') \\  
-1 & -a_R(\varepsilon') \end{array} \right) \, ,  
\label{d10} \end{equation} 
and also show explicitly that the advanced functions $G_A$ are related 
to $G_R$ as follows: 
\begin{equation} 
G_A(\varepsilon,\varepsilon',\mbox{sign}(p_x)) = 
- G^*_R(\varepsilon,\varepsilon',-\mbox{sign}(p_x)) \, . 
\label{gar} \end{equation} 

The current in the constriction depends not only on the density of 
states but also on the occupation of these states. The information about 
occupation probabilities is contained in the Keldysh Green's function 
$G_K (\varepsilon,\varepsilon')$ which satisfy the following equation 
\cite{lo,zai1}: 
\begin{equation}
i \mbox{sign}(p_x)v_F \frac{\partial G_K }{\partial x}=
H_RG_K+H_KG_A-G_RH_K-G_KH_A\, , 
\label{d11} \end{equation} 
where $H_{R,A}$ are the same matrices which determine the 
evolution of $G_{R,A}$ in eq.\ (\ref{d2}), $H_K=H_Rn-nH_A$,  
and $n$ is the equilibrium quasiparticle distribution, 
$n(\epsilon,\epsilon')= \tanh (\epsilon /2T) \delta(\epsilon-\epsilon')$.

Equation (\ref{d2}) implies that solution of eq.\ (\ref{d11}) can be 
written as 
\begin{equation} 
G_K=G_Rn-nG_A+G_H\, , 
\label{gk} \end{equation} 
where $G_H$ satisfies the homogeneous equation:
\begin{equation}
i\mbox{sign}(p_x)v_F\frac{\partial G_H}{\partial x}=H_RG_H-G_HH_A\, . 
\label{18} \end{equation}
Calculation of $G_H$ follows closely the one for $G_{R,A}$ that was 
described above. Diagonalizing $H_{R,A}$ in eq.\ (\ref{18}) by 
transformation (\ref{d3}) 
and imposing the condition that $G_H$ decays inside the electrodes, 
we see that for $p_x>0$ and $x=0$,  
$\tilde{G}_{H1,2}\propto (1\mp \sigma_z)$ in the first and second 
electrode respectively. From this we find that  
\[ G_{H1}(\varepsilon,\varepsilon') =  U_R^{-1}\tilde{G}_{H1} U_A = \]
\vspace*{-3ex} 
\begin{equation} 
= h_1(\varepsilon,\varepsilon')  \left( \begin{array}{cc} 
1 & a_A(\varepsilon') \\ -a_R(\varepsilon) & 
-a_R(\varepsilon) a_A(\varepsilon') \end{array} \right) \, , 
\label{gh} \end{equation}
\[  
G_{H2}(\varepsilon,\varepsilon') =  U_R^{-1}\tilde{G}_{H2} U_A = \] 
\vspace*{-3ex} 
\begin{equation} 
= h_2(\varepsilon,\varepsilon')  \left( \begin{array}{cc} 
-a_R(\varepsilon) a_A(\varepsilon') & -a_R(\varepsilon) 
\\ 1 & a_A(\varepsilon') \end{array} \right) \, . 
\label{gh2} \end{equation} 

Next we impose the continuity condition $G_{K1}=G_{K2}$ in 
the constriction, and eliminating $h_2$ from this condition obtain 
the recurrence relation for $h_1$. This recurrence relation can be 
solved similarly to the recurrence relation (\ref{d8}) and gives: 
\[ h_1(\varepsilon,\varepsilon')=  2\sum_{m=0}^{\infty} 
\left[ \sum_{n=0}^{\infty} \delta(\varepsilon-\varepsilon'+2un) 
\prod_{j=1}^{2n+m} a_R(\varepsilon+ju)  \right. \] 
\vspace*{-2ex} 
\[ \prod_{j=1}^{m} a_A(\varepsilon' +ju) n^{(-)} (\varepsilon'+mu) 
+ \sum_{n=1}^{\infty} \delta(\varepsilon-\varepsilon'-2un) \] 
\vspace*{-2ex} 
\begin{equation} 
\left. \prod_{j=1}^{m} a_R(\varepsilon+ju)  \prod_{j=1}^{2n+m} 
a_A(\varepsilon' +ju) n^{(-)} (\varepsilon+mu) \right] \, ,  
\label{h1} \end{equation} 
where $n^{(-)} (\varepsilon)\equiv n(\varepsilon+u)-n(\varepsilon)$, 
and we used the convention that $\prod_{j=1}^{0}(...)=1$.  
Equations (\ref{gh}), (\ref{h1}) and (\ref{gk}) determine the Keldysh 
Green's function in the constriction for $p_x>0$. Following the same 
steps we can also find $G_K$ for $p_x<0$. 

The time-dependent current $I(t)$ in the constriction can be expressed 
in terms of $G_K$ as 
\[ I(t) =\frac{e}{8\pi\hbar} \int d\epsilon d\epsilon' 
\mbox{Tr} \left[ \sigma_z \left( G_K^{(p_x>0)}(\epsilon,\epsilon') 
- \right. \right.\] 
\vspace*{-3ex} 
\begin{equation} 
\left. \left. - G_K^{(p_x<0)}(\epsilon,\epsilon') \right) \right] 
\exp^{i (\epsilon-\epsilon')t/\hbar}  \, .
\label{curr} \end{equation}
For $G_K$ found above (and given by the eqs.\ (\ref{gk}), (\ref{gh}), 
(\ref{h1}), (\ref{d9}) and (\ref{gar})), eq.\ (\ref{curr}) shows that 
the current is a combination of harmonics of the Josephson frequency 
$\omega_J = 2eV/\hbar$, and the amplitude of the $k$th harmonics is
\cite{we2}: 
\[ I_k =\frac{e}{\pi \hbar} \left[ eV \delta_{k0} - \int d\epsilon 
\tanh( \frac{\epsilon}{2T} ) (1-\mid a_R(\epsilon )\mid^2) 
\right. \]
\begin{equation}
\sum_{m=0}^{\infty} \prod_{j=1}^m \mid a_R(\epsilon +jeV)\mid^2 
\left. \prod_{j=m+1}^{m+2k} a_R(\epsilon +j eV) \right]  \, .
\label{20} \end{equation}   
Transformations bringing expression for the current into this 
form used the relation $a_R(-\epsilon)=-a_R^*(\epsilon)$. This relation 
holds due to assumed electron-hole symmetry of the superconducting 
electrodes of the constriction. 

\begin{figure}[htb]
\setlength{\unitlength}{1.0in}
\begin{picture}(2.0,4.7) 
\put(0.25,0.1){\epsfxsize=3.0in\epsfysize=4.5in \epsfbox{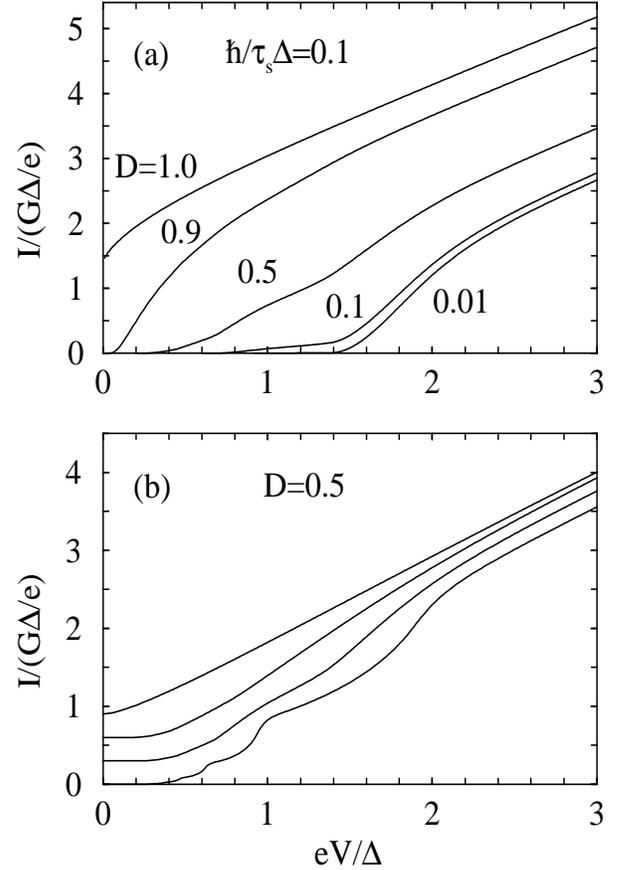}}
\end{picture}
\caption{DC $IV$ characteristics of a single-mode constriction  
between two superconductors with paramagnetic impurities for  
several values of (a) constriction transparency $D$, and (b) 
spin-flip scattering rate $1/\tau_s$ in the superconductors. $G$ in
normalization of the current is the normal-state conductance of the 
constriction $G=e^2D/\pi \hbar$. The 
curves in (b) are shifted for clarity and correspond to (from bottom 
to top) $\hbar/\tau_s\Delta=0.01,\, 0.1,\, 0.3,\, 1.0$. For discussion 
see text.}
\end{figure}

Comparison of eq.\ (\ref{20}) for the Fourier harmonics of the 
current in a single-mode constriction with the $D=1$ version of eq.\ 
(\ref{ccc9}) obtained from the BdG equations shows that these two 
equations coincide if we identify $a_R$ with the amplitude of the 
Andreev reflection. To see this equivalence we note that at $D=1$ 
the recurrence relations (\ref{c9}) and (\ref{c10}) give that 
$B_n\equiv 0$ for all $n$, $A_n=0$ for $n\leq 0$, and 
$A_n=\prod_{j=1}^{2n-1}a_j$ for $n>0$. With these wavefunction 
amplitudes eq.\ (\ref{ccc9}) indeed reduces to eq.\ (\ref{20}).  
This means that the only role of the complex  
internal structure of the superconducting electrodes (including 
possible inelastic processes) in determining the ac Josephson current 
in a short ballistic constriction is to modify the amplitude of the 
Andreev reflection which is in general given by eq.\ (\ref{d4}). 
Starting from the Zaitsev's solution of the constriction problem 
in the matrix form \cite{zai2}, one can show \cite{we4} that this 
conclusion is valid for arbitrary $D$. The Fourier components 
of the time-dependent current in the constriction are determined by  
eq.\ (\ref{ccc9}) and recurrence relations (\ref{c9}) and (\ref{c10}) 
for general microscopic structure of the superconducting electrodes, 
if  the amplitude of Andreev reflection $a_R$ is defined 
by eq.\ (\ref{d4}) \cite{mr3}.

To illustrate this approach, we calculated the $IV$ characteristic of 
a short constriction between two superconductors with paramagnetic 
impurities. In this case, the BCS singularity in the density of states 
at the gap edge is broadened by the spin-flip scattering and the 
magnitude of the  gap itself is suppressed. These two effects are reflected 
in the energy dependence of the Andreev reflection amplitude 
$a_R(\varepsilon)$ which can be written in a form similar to eq.\ 
(\ref{aa}): 
\begin{equation}
a_R(\varepsilon) = u(\varepsilon)-(u(\varepsilon)^2- 1)^{1/2} \, , 
\label{ap} \end{equation} 
where $u(\varepsilon)$ is given by the equation \cite{maki}: 
\begin{equation}
\frac{\varepsilon}{\Delta} = u \left( 1-\frac{\hbar/\tau_s\Delta }{
(1-u^2)^{1/2}} \right) \, .
\label{uu} \end{equation} 
Here $1/\tau_s$ is the rate of the spin-flip scattering and $\Delta$ 
is the energy gap in absence of this scattering. Equations (\ref{ap}) 
and (\ref{uu}) allow for a simple numerical determination of 
$a_R(\varepsilon)$. DC $IV$ characteristics for zero temperature and 
several scattering rates $1/\tau_s$ and transparencies $D$ calculated 
with the $a_R(\varepsilon)$ from eqs.\ (\ref{ap}) and (\ref{uu}) 
are plotted in Fig.\ 3. Figure 3a shows how the 
$IV$ characteristic evolves as a function of the constriction 
transparency $D$ at weak spin-flip scattering. We see that even weak    
scattering results in effective broadening of all current 
singularities. Due to suppression of the energy gap in the electrodes, 
the spin-flip scattering also gives rise to suppression of the zero-bias 
current jump at $D=1$ and suppression of the excess current. Such a 
gap suppression becomes 
progressively more pronounced at larger scattering rates (Fig.\ 3b), 
until the gap disappears completely at $\hbar/\tau_s =\Delta$. In 
the gapless regime, the $IV$ characteristic is practically linear.  

\section{Noise} 

The method developed in the previous Section allows us to calculate 
not only the average current in the constriction, but also 
fluctuations of the current around the average. The spectral density 
of the current fluctuations $S_I(\omega)$ in presence of the 
time-dependent average current is defined as 
\begin{equation} 
S_I(\omega) =\frac{1}{2\pi} \int d\tau e^{i\omega\tau}\
\overline{K_I(t,t+\tau)}  \, , 
\label{si} \end{equation} 
\[K_I(t,t+\tau)\equiv \frac{1}{2} \langle I(t)I(t+\tau)+I(t+\tau)I(t) 
\rangle  - \langle I(t)\rangle \langle I(t+\tau) \rangle \, , \] 
where the bar over $K_I(t,t+\tau)$ denotes averaging over the time $t$.
In this Section we discuss only the single-mode ballistic constriction. 
For such a constriction, the correlation function $K_I$ can be expressed 
in terms of the quasiclassical Green's functions \cite{khlus}: 
\[ K_I(t,t+\tau) = -\frac{e^2}{8} \sum_{\pm} \mbox{Tr} [ 
G_{\pm}^>(t,t+\tau)\sigma_z G_{\pm}^<(t+\tau,t)\sigma_z + \] 

\vspace*{-2ex}

\[ + G_{\pm}^<(t,t+\tau)\sigma_z G_{\pm}^>(t+\tau,t)\sigma_z ] \, ,  \] 
where $\sum_{\pm}$ is the sum over the two directions of propagation 
in the constriction. The Green's functions $G^{>,<}$ are related to 
the functions $G_{R,A,K}$ discussed in the previous section:  
\begin{equation} 
G^{>,<} =\frac{1}{2} [ G_K \pm (G_R-G_A)] \, . 
\label{ggl} \end{equation} 
The functions $G(\varepsilon,\varepsilon')$ are the Fourier transforms 
of the $G(t,t')$. 

If we use in eq.\ (\ref{ggl}) the Green's functions found in the 
previous section, we see that $G^{>,<}$ can be written as 
\begin{equation}
G^{>,<}_{\pm} (\varepsilon,\varepsilon') =\sum_n G^{>,<}_{n,\pm} 
(\varepsilon+neV) \delta(\varepsilon-\varepsilon'+2neV) \, , 
\label{qn} \end{equation} 
and the spectral density (\ref{si}) assumes the form: 
\begin{equation} 
S_I (\omega ) = -\frac{e^2}{32 \pi^2 \hbar} \sum_{n,\, \pm, \pm \omega} 
\int d\varepsilon \mbox{Tr} [ G_{n,\pm}^>(\varepsilon ) \sigma_z 
G_{-n,\pm}^< (\varepsilon \pm \hbar \omega ) \sigma_z ] \, .  
\label{si2} \end{equation}
Combinig this equation with the equations for the Green's function 
$G^{>,<}$ (eqs.\ (\ref{ggl}) and (\ref{qn}) of this Section and 
eqs.\ (\ref{gk}), (\ref{gh}), (\ref{h1}), (\ref{d9}) and (\ref{gar}) of 
the previous Section) we arrive at the final result for $S_I(\omega )$ 
\cite{we3}:

\[ S_I (\omega ) = \frac{e^2}{2 \pi^2 \hbar} \sum_{ \pm \omega} 
\int d\varepsilon F(\varepsilon)(1-F(\varepsilon \pm \hbar \omega)) \] 

\vspace*{-2ex}

\begin{equation} 
[1+2\mbox{Re} \sum_{k=1}^{\infty} \prod_{j=1}^k a_R(\varepsilon+jeV)
a_A(\varepsilon+jeV\pm \hbar \omega) ]\, .  
\label{si3} \end{equation}
Function $F$ here has the meaning of non-equilibrium distribution of 
quasiparticles in the constriction: 
\[ F(\varepsilon)= f(\varepsilon)+ \sum_{n=0}^{\infty} \prod_{m=0}^n \mid 
a_R(\varepsilon-meV) \mid^2 \] 

\vspace*{-2ex} 

\begin{equation}
[f(\varepsilon-(n+1)eV) - f(\varepsilon-neV)] \, ,
\label{dis} \end{equation}
where $f(\varepsilon)$ is the Fermi distribution function. 

\begin{figure}[htb]
\setlength{\unitlength}{1.0in}
\begin{picture}(2.0,2.5) 
\put(0.35,0.15){\epsfxsize=2.7in\epsfysize=2.6in \epsfbox{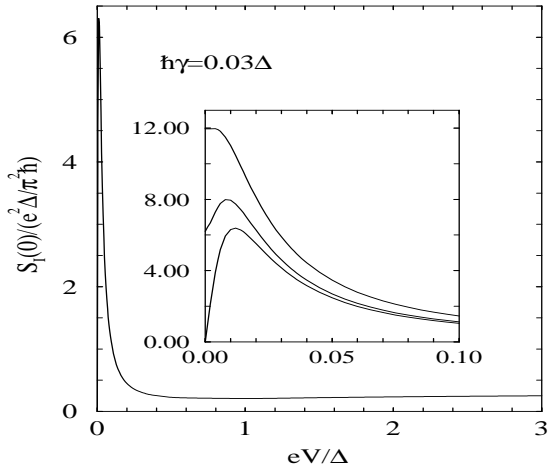}}
\end{picture}
\caption{Zero-frequency spectral density of fluctuations of the current 
in a single-mode ballistic constrictions at zero temperature as a 
function of the bias voltage. The peak at low voltages is due to 
multiple Andreev reflections. The inset shows a blowup of the peak at 
several temperatures: (from  bottom to top) $T/\Delta= 0,\, 1,\, 2$.}  

\end{figure}

The spectral density of current fluctuations defined by eqs.\ 
(\ref{si3}) and (\ref{dis}) depends strongly on the behavior 
of the Andreev reflection amplitude $a_R(\varepsilon)$ as a 
function of energy $\varepsilon$. Figure 4 shows the zero-frequency 
spectral density $S_I(0)$ calculated numerically for the 
superconductors with small quasiparticle inelastic scattering rate
$\gamma$, when $a_R(\varepsilon)$ is given by eq.\ (\ref{ap}) with 
\[ u(\varepsilon)=\frac{\varepsilon +i\hbar \gamma/2}{\Delta} \, .\] 
One can see from Fig.\ 4 that in this case the main feature of the 
current noise is the peak of $S_I(0)$ at low voltages. The height of 
the peak is much larger than the classical shot noise result $eI/2\pi$ 
which in our case would be on the order of $e^2\Delta/2\pi^2\hbar$. 
To understand the origin of this large noise  we evaluate eqs.\ 
(\ref{si3}) and (\ref{dis}) analytically in the limit $V\ll \Delta/e$. 
Expanding the 
amplitude $a_R(\epsilon)$ of Andreev reflection in small relaxation 
rate $\gamma$ and replacing the sums with the integrals we obtain the 
spectral density of current 
fluctuations at low frequencies, $\omega \ll \Delta/\hbar$: 
\[ S_I (\omega ) =  \frac{e}{\pi^2 \hbar V} \sum_{\pm \omega} 
\int_{- \Delta}^{\Delta} d\varepsilon F(\varepsilon)(1- F(\varepsilon 
\pm\hbar \omega)) \] 

\vspace*{-2ex} 

\[ \int^{\Delta}_{\varepsilon }d\varepsilon' \exp \{ - 
\int_{\varepsilon }^{\varepsilon'}\frac{d\nu \hbar \gamma(\nu)}{eV 
\sqrt{\Delta^2-\nu^2}} \} \]

\vspace*{-2ex} 

\begin{equation}
\cos ( \frac{\hbar \omega}{eV} [\arccos (\frac{\varepsilon'}{\Delta}) - 
\arccos (\frac{\varepsilon}{\Delta}) ] ) \, , 
\label{low} \end{equation}
where the quasiparticle distribution function reduces to  
\begin{equation}
F(\varepsilon) = f(\varepsilon ) - \int_{-\Delta}^{\varepsilon }
d\varepsilon' 
\frac{\partial f}{\partial \varepsilon'}\exp \{ -\int_{ \varepsilon' }^{ 
\varepsilon }\frac{ d \nu \hbar \gamma (\nu)}{eV \sqrt{\Delta^2-\nu^2}} 
\} \, .  \label{dis2} \end{equation} 

For ideal BCS superconductors $\gamma \rightarrow 0$ and eq.\ (\ref{dis2}) 
immediately gives that $F(\varepsilon)=f(\Delta)$. Then the spectral 
density (\ref{low}) can be found explicitly: 
\begin{equation} 
S_I(\omega) = \frac{e\Delta^2 }{2\pi^2 \hbar \cosh^2 (\Delta/2T) V}\,
\frac{1+ \cos (\pi \hbar \omega/eV)}{(1-(\hbar \omega /eV )^2)^2}\, . 
\label{fin} \end{equation} 
This equation shows that the noise diverges at $V\rightarrow 0$ as 
$1/V$. At finite relaxation rate $\gamma$ this divergence saturates at 
$V\simeq \hbar\gamma/e$ and gives the low-voltage peak in Fig.\ 4. 
The origin of this unusual voltage dependence of the noise is the process 
of multiple Andreev reflections. Each quasiparticle entering 
the constriction with the energy equal to one of the gap edges 
generates an avalanche of Andreev reflections before it can escape 
out of the constriction by climbing up or down in energy to the 
opposite edge of the energy gap. The number of generated Andreev 
reflections is $2\Delta/eV$, so that each quasiparticle causes a 
coherent transfer through the constriction of a charge quantum of 
magnitude $2\Delta /V$. For small voltages $V\ll \Delta/e$ this charge 
is much larger than the charge of an individual Cooper pair. In this way 
the randomness of the quasiparticle scattering (quasiparticles get 
inside the energy gap with probability $f(-\Delta)$ from one electrode 
and with probability $f(\Delta)$ from the opposite electrode) is 
amplified. Therefore, the noise described by eq.\ (\ref{fin}) can 
be interpreted as the shot noise of these large charge quanta, and  
we see that the process of multiple Andreev reflections dominates the 
current noise as well as the average current in short ballistic 
superconducting constrictions. 

This energy-domain picture with cycles of multiple Andreev reflections, 
together with quantitative eqs.\ (\ref{low}) and (\ref{dis2}), has the 
``dual'' time-domain formulation in terms of evolution of occupation 
probabilities of the two quasi-stationary Andreev-bound states localized 
in the constriction that carry dc Josephson current \cite{me}. An advantage of 
the time-domain formulation 
is that it can be generalized in a straightforward way to 
situations with the  time-dependent bias voltage. It is also convenient 
in establishing explicit relation between the ac, finite-voltage regime 
and dc regime in dynamics of superconducting constrictions. In particular 
it shows how the finite-voltage current noise discussed 
in this Section goes over into large equilibrium supercurrent noise at
vanishing bias voltage \cite{we3,mr}.

\section{Conclusions and unsolved problems}

We have seen in the preseeding Sections that all electron transport 
properties of short superconducting constrictions are determined by the
interplay of quasiparticle scattering inside the constriction and Andreev 
reflection at the interfaces between the constriction and superconducting 
electrodes. Quantitatively, the scattering process at finite bias voltages  
across the constriction is described by a set of recurrence relations for 
the amplitudes of quasiparticle wave functions at energies $\varepsilon_n$ 
shifted by integer number $n$ of quanta of the Josephson oscillation, 
$\varepsilon_n=
\varepsilon+ 2eVn, \; n=0,\pm 1, ...$. The recurrence relations are valid 
for general scattering properties of the constriction that are characterized 
by the set of transmission eigenvalues $D_m$, and general microscopic 
structure of the superconducting electrodes characterized by the energy
dependence of the Andreev reflection amplitude $a(\varepsilon)$.  

In this Chapter, we discussed the constrictions between identical 
$s$-wave superconductors. The recurrence relations derived for this 
situation can be generalized to 
the case of two different superconductors (see the Chapter by P. Bagwell 
{\em et al.} in this volume) and $d$-wave superconductors \cite{tan,hurd}. 
An interesting possible direction for 
generalization of the recurrence relations is the case of transmission 
coefficient $D(\varepsilon)$ that is energy-dependent on the scale of
superconducting energy gap $\Delta$. Such a generalization would 
allow, for example, a systematic study of the ac Josephson effect 
through a resonant level \cite{mr2,joh} or thick tunnel barrier. Another 
important unsolved problem is the extension of the recurrence relations 
approach to the constrictions with the spin-dependent scattering, for 
example, Josephson junctions with ferromagnetic interlayer -- see the 
contribution of G. Arnold {\em et al.} to this volume. 

The basic simplification which allows to describe quantitatively such 
a variety of different situations is the approximation of short 
constriction $d\ll \xi$. A very important open problem is the lifting   
of this restriction. General solution of this problem for junctions with 
$d\simeq \xi$ appears very difficult due to the necessity of 
self-consistent determination of the distribution of the pair potential 
$\Delta(x)$ and 
electrostatic potential $\varphi(x)$. As a first step towards 
such a solution, it would be interesting to apply the MAR approach 
to the opposite limit of long junctions $d\gg \xi$. Long disordered 
SNS junctions have been attracting a lot of attention recently -- 
see, e.g., \cite{sns1,sns3,sns5} and references therein, but the interest 
has been focused mainly on their properties close to equilibrium. 
MAR approach can be advantageous in studying strongly non-equilibrium
situations. Another 
challenging unsolved problem is application of the results presented 
in this chapter to the description of Coulomb blockade in Josephson 
junctions with large transparency. The Coulomb blockade regime is 
characterized by quantum dynamics of the Josephson phase difference 
$\varphi$ \cite{set,sch}, in contrast to the classical dynamics 
of $\varphi$ assumed in this Chapter.    

In summary, one can expect further exciting new developments and rapid 
progress in understanding of high-transparency Josephson junctions.

\end{document}